\documentclass[fleqn,usenatbib]{mnras}

\usepackage{newtxtext,newtxmath}
\usepackage{halloweenmath}

\usepackage[T1]{fontenc}

\DeclareRobustCommand{\VAN}[3]{#2}
\let\VANthebibliography\thebibliography
\def\thebibliography{\DeclareRobustCommand{\VAN}[3]{##3}\VANthebibliography}

\usepackage{graphicx}	% Including figure files
\usepackage{amsmath}	% Advanced maths commands
\usepackage{lipsum}  

\title[Mixed Dark Matter]{Pushing the Limits of Detectability: Mixed Dark Matter from Strong Gravitational Lenses}

\author[R. E. Keeley et al.]{
Ryan E. Keeley,$^{1}$\thanks{E-mail: rkeeley@ucmerced.edu}
Anna M. Nierenberg,$^{1}$
Daniel Gilman$^{2}$
Simon Birrer$^{3,4}$
Andrew Benson$^{5}$
Tommaso Treu$^6$
\\
$^{1}$Department of Physics, University of California Merced, 5200 North Lake Road, Merced, CA 95343, USA\\
$^{2}$Department of Astronomy and Astrophysics, University of Toronto, Toronto, ON, M5S 3H4, Canada\\
$^{3}$Kavli Institute for Particle Astrophysics and Cosmology, Department of Physics, Stanford University, Stanford, CA, USA\\
$^{4}$SLAC National Accelerator Laboratory, Menlo Park, CA, USA\\
$^{5}$Carnegie Institution for Science, 813 Santa Barbara Street, Pasadena, CA 91101, USA\\
$^{6}$Department of Physics and Astronomy, University of California, Los Angeles, CA, 90095
}

\date{Accepted XXX. Received YYY; in original form ZZZ}

\pubyear{2023}

\begin{document}
\label{firstpage}
\pagerange{\pageref{firstpage}--\pageref{lastpage}}
\maketitle

\begin{abstract}
One of the frontiers for advancing what is known about dark matter lies in using strong gravitational lenses to characterize the population of the smallest dark matter halos. 
There is a large volume of information in strong gravitational lens images---the question we seek to answer is to what extent we can refine this information.
To this end, we forecast the detectability of a mixed warm and cold dark matter scenario using the anomalous flux ratio method from strong gravitational lensed images.
The halo mass function of the mixed dark matter scenario is suppressed relative to cold dark matter but still predicts numerous low-mass dark matter halos relative to warm dark matter.
Since the strong lensing signal receives a contribution from a range of dark matter halo masses and since the signal is sensitive to the specific configuration of dark matter halos, not just the halo mass function, degeneracies between different forms of suppression in the halo mass function, relative to cold dark matter, can arise.
We find that, with a set of lenses with different configurations of the main deflector and hence different sensitivities to different mass ranges of the halo mass function, the different forms of suppression of the halo mass function between the warm dark matter model and the mixed dark matter model can be distinguished with $40$ lenses with Bayesian odds of 30:1.
\end{abstract}


%%%%%%%%%%%%%%%%%%%%%%%%%%%%%%%%%%%%%%%%%%%%%%%%%%

%%%%%%%%%%%%%%%%% BODY OF PAPER %%%%%%%%%%%%%%%%%%

\section{Introduction}

$\Lambda$CDM, $\Lambda$ for a cosmological constant type dark energy, and CDM for cold dark matter (DM), has maintained its status as the standard model of cosmology over the past generation of cosmological observations.
With just a few parameters, it can explain most of the Universe.
The broader goal of much of cosmology is to identify the nature of these components of $\Lambda$CDM, in particular for astroparticle physics, the nature of DM. One strategy to identify the nature of DM is to look for any Standard Model (SM) DM annihilation products (indirect detection)~\citep{Gaskins:2016cha}, collisions of DM with SM particles (direct detection)~\citep{Schumann:2019eaa}, or to produce DM from SM particles at colliders~\citep{Kahlhoefer:2017dnp}.

Examples of specific DM models include the weakly interacting massive particles (WIMP) and sterile neutrinos, with all of their various production mechanisms \citep{Dodelson:1993je,Shi:1998km,Abazajian:2001nj,Kusenko:2009up,Abazajian:2017tcc,Abazajian:2019ejt}.
However, the simplest indirect detection of WIMP annihilation from the largest, closest DM source to Earth, the Galactic center, is ruled out~\citep{Abazajian:2020tww}.  Thus it is prudent to look at other probes for hints of DM's particle nature, though measuring the distribution of DM (e.g. halo-mass function) is interesting regardless of any results from indirect detection.

For instance, rather than probing DM's interactions with the SM, one can probe DM's phenomenological properties (i.e. whether there is a suppression in the amount of clustering relative to CDM, or whether DM interacts with itself) via its gravitational interactions. The CDM paradigm, which posits DM is collisionless as well as cold, predicts the existence of collapsed DM halos down to very small halo masses (equivalently very short length scales)~\citep{Metcalf:2001ap,Dutton:2014xda,Despali:2015yla,Angulo:2016qof}. Thus, measuring the distribution of DM on small scales can serve as a useful determination of DM phenomenology and can serve as a hint towards identifying the particle nature of DM. Importantly, the distribution of DM throughout the Universe can be probed by only characterizing its gravitational interactions, such as with strong gravitational lensing~\citep{Mao:1997ek,Dalal:2001fq,Metcalf:2001ap,Moustakas:2002iz,Chen:2003uu,Amara:2004dr,Metcalf:2004eh,Miranda:2007rb,Minezaki:2009ek}.

The missing satellite problem is motivation for thinking that warm DM (WDM) might more accurately describe the Universe than CDM~\citep{Viel:2013fqw}. The smallest DM halos, that $\Lambda$CDM predicts in abundance, were not observed in the Local Group or cosmologically, such that it was once thought to be a possibility that the halo mass function was suppressed on the dwarf scale~\citep{Nierenberg:2016nri,Robles:2018fur}. However, the most recent generations of telescopes have detected populations of ultra-faint dwarf galaxies, which have, for certain assumptions about star formation and completeness corrections, constrained deviations between WDM and CDM subhalo mass functions to be only on sub-galactic scales.
Further, some simulations, along with specific assumptions about star formation, baryonic feedback and tidal stripping, show that subhalos are destroyed more efficiently than originally thought from DM only simulations~\citep{Kim:2017iwr,2021arXiv210609050K}.

Beyond either the CDM or WDM paradigms, the cosmological DM could be composed of multiple particles with different phenomenological properties~\citep{Boyarsky:2008xj,2012JCAP...10..047A,Kamada:2016vsc,Parimbelli:2021mtp,Vogt:2022bwy}.  Specifically, we investigate a case where half of the DM is composed of CDM particles and half is WDM particles. We refer to this case as mixed DM (MixDM).
There are a wide variety of plausible scenarios in which MixDM could be realized.  For instance, many models have three generations of sterile neutrinos with different masses from keV to GeV~\citep{Kusenko:2009up,Patwardhan:2015kga,Abazajian:2017tcc}. With different masses and potentially different resonances in their production mechanism, such multiple generations of sterile neutrinos could have a complicated transfer function compared to CDM~\citep{Abazajian:2019ejt,Vogel:2022odl}. Another possibility is that WIMPs could play the role of CDM and axions could play the role of fuzzy DM (as the component that has less power on small scales)~\citep{2020PrPNP.11303787N}.  We are not trying to make the case that MixDM should be a priori expected, but to present a not-unreasonable model that offers a test for what should be detectable with upcoming observations of strong lens systems with JWST.

In strong gravitational lensing, the light from a source is deflected by the combined gravitational potential of a main deflector lens and all of the subhalos and line-of-sight halos along the trajectory the light follows to create multiple images of the source.  In certain configurations of the lens, four distinct images will be created. 
Analyzing the fluxes of these images can allow us to make inferences about the mass function of low mass DM halos~\citep{Mao:1997ek,Dalal:2001fq,Metcalf:2001ap,Moustakas:2002iz,Chen:2003uu,Amara:2004dr,Metcalf:2004eh,Miranda:2007rb,Minezaki:2009ek,Gilman:2016uit,Gilman:2017voy,Gilman:2019bdm,2022MNRAS.517.1867L}.

Further, since strong gravitational lensing can probe completely dark halos, it can probe the physics of the least massive halos at the smallest scales, both in the lens and along the line of sight, which is where deviations from CDM are expected to be found. Because they are dark, any potential signal of DM physics would not be confused for new baryonic/stellar physics. Low-mass DM halos are hard to probe since they are not efficient at forming galaxies.
However, these low-mass DM halos can be ``seen'' via their gravitational interactions, such as in strong gravitational lensing of quasars by galaxies. New DM physics from WDM, self-interacting DM (SIDM), collisional, and fuzzy DM predict novel configurations and distributions of low-mass DM halos that could explain the small-scale structure of DM halos.  Thus, characterizing the distribution and profiles of these lowest mass DM halos with strong gravitational lenses could provide evidence for novel DM physics.

Strong lens systems can be used not just to infer the abundance of DM halos but also their concentrations~\citep{Gilman:2019bdm}.  Indeed, recent works have found subhalos that have a concentration much higher than would be expected in $\Lambda$CDM and may point towards SIDM~\citep{Andrade:2019wzn,Andrade:2020lqq,Minor:2020bmp,Minor:2020hic,Gilman:2022ida}.

There are two techniques that are commonly employed that use strong gravitational lenses to infer the properties of DM: the flux ratio anomaly method and gravitational imaging. 
The key difference between these methods is the size of the source observed. The flux ratio anomaly method, which we employ in this work, uses sources of sufficiently small size, such that the lens will create four distinct point sources\citep{Dalal:2001fq,Gilman:2019vca,Gilman:2019bdm,2020MNRAS.492.3047H,Gilman:2021sdr,Gilman:2022ida}. In contrast, gravitational imaging uses larger sources, such that the lens will create extended arcs~\citep{2018MNRAS.481.3661V,2021MNRAS.506.5848E}.  
The idea with the flux ratio anomaly method is to examine perturbations in the fluxes of quadruply imaged quasars, relative to the predictions of the main deflector.  The anomalous flux ratio method works by using the image positions to constrain the smooth mass distribution of the main deflector lens and predict values for the observed flux ratios. Any additional perturbations from subhalos and line of sight halos will affect the flux ratios. These perturbations amount to detections of a population of DM halos and the modeling of these perturbations, in a statistical sense, can yield information about the statistical properties of the population of DM halos, and thus the physics that generated them. Existing constraints using this method place the mass of a thermally produced WDM relic at $m_{\rm WDM}>9.7$ keV~\citep{Nadler:2021dft}. There is a lot of information in these quad lens system and so we seek to understand how far we can refine that information.

In this paper, we investigate the potential for strong lens systems, as measured by upcoming JWST observations, to detect novel DM physics beyond either the CDM and WDM paradigms, e.g., a mixed DM (MixDM) model.  We elaborate on the details of this model in Sec.~\ref{sec:Mix}. We further discuss the details of the flux ratio anomaly method from strong lens systems in Sec.~\ref{sec:SLDetails} and the details of our statistical methods in Sec.~\ref{sec:Statistics}.  We present the results of our forecasts in sec.~\ref{sec:Results} and conclude in Sec.~\ref{sec:Conclusions}.

\section{Mixed Dark Matter}\label{sec:Mix}

One way to extend the CDM paradigm is to allow the DM particle to be slightly warm.  This WDM particle would free stream a non-negligible distance over the age of the Universe and the WDM particle would therefore diffuse out of small overdensities in the matter field and delay the collapse and growth of DM halos. Any amount of free streaming of particle dark matter can wash out structure on scales below the free-streaming length~\citep{1983ApJ...274..443B,2004MNRAS.353L..23G}. Lighter particles free stream for longer times and thus will wipe out more structure. For thermally-produced relics, WDM composes a subclass of sub-GeV DM candidates. It is this special case for which constraints on a WDM mass are often quoted, but the power of arguments based on structure formation allows one to recast an inference of the free-streaming length in the context of any dark matter model with a cosmologically-relevant free-streaming length, such as sterile neutrinos~\citep{Zelko:2022tgf}.

\begin{figure}
    \centering
    \includegraphics[width=\columnwidth]{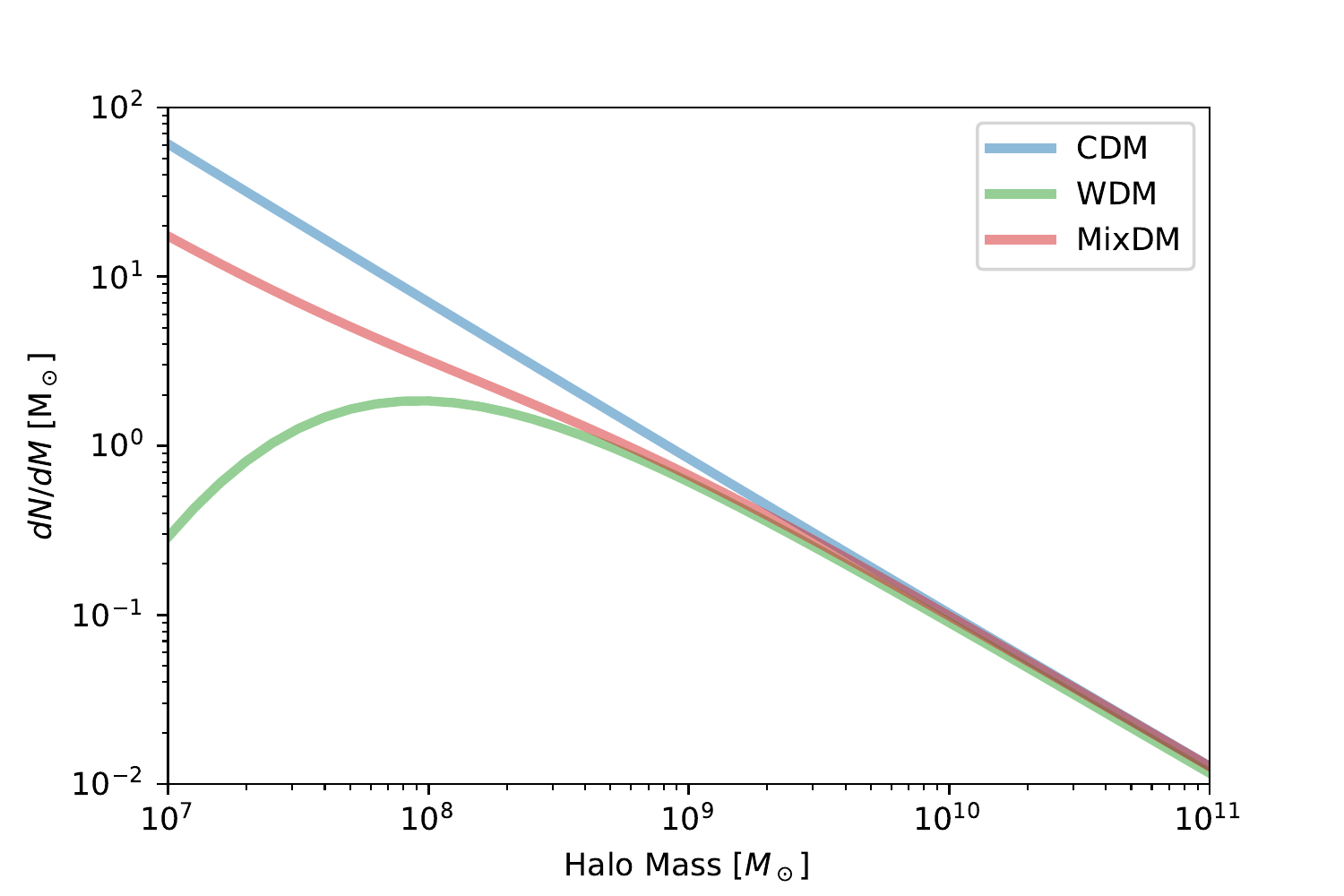}
    \caption{Example halo mass functions for CDM (blue), WDM (green), and MixDM (red) cases where the warm component of the MixDM case has the same mass as the WDM case, specifically, that the $M_{\rm hm} = 10^{8.5}$ M$_\odot$ for both models.}
    \label{fig:mixed_hmf}
\end{figure}

\begin{figure*}
    \centering
    \includegraphics[width=\textwidth]{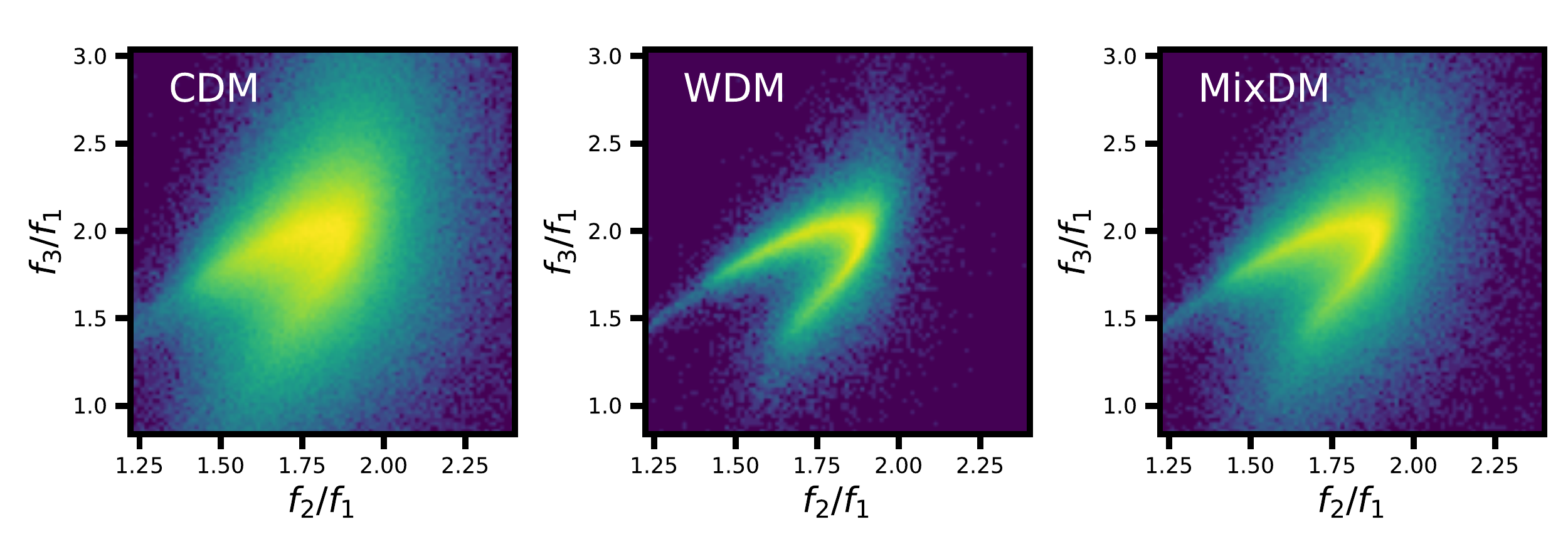}
    \caption{Example 2D histograms of flux ratios ($f_i$ are the flux ratios of individual images) for CDM, WDM, and MixDM cases. The width of the distributions correspond to the statistical signal that we seek to use to differentiate these models. }
    \label{fig:ex_ratios}
\end{figure*}

Examples of a WDM candidate include sterile neutrinos, axinos, gravitinos or any light thermally produced relic~\citep{1994PhRvL..72...17D,1996ApJ...458....1C,Abazajian:2017tcc,Vogel:2022odl}. The microphysics of specific WDM models can influence greatly the mapping between particle properties, such as the mass, and astrophysical observables, such as the power spectrum~\citep{Abazajian:2017tcc,Vogel:2022odl}.  Typically, however, when lower limits on DM masses are quoted, it is assumed the WDM particle followed a thermal distribution at early times. This is important to keep in mind, since if, for example, a sterile neutrino was resonantly produced with lower momentum, compared to a generic thermal relic of the same mass, then it would appear colder~\citep{1999PhRvL..82.2832S,Abazajian:2017tcc}. Such a model could include a ``chilly'' sterile neutrino. Thus certain models of sterile neutrino production can evade existing bounds on the WDM mass yet still explain potential signals like the 3.55 keV line which would be explained by the decay of a 7.1 keV sterile neutrino~\citep{Abazajian:2017tcc}.

Similarly, the half-mode mass ($M_{\rm hm}$), the mass of the DM halo that corresponds to a decrease in the WDM transfer function by a factor of one half relative to the CDM transfer function, is a useful parameter that summarizes the suppression of structure at small scales.  This is a characteristic scale at which a suppression in small-scale structure should become observable.  We can probe half-mode masses down to $10^{6.5} \mathrm{M}_\odot$ with strong lensing observed with the JWST. In an upcoming JWST program (GO-2046, PI Nierenberg), we will measure flux ratios with $\sim3\%$ relative precision, which, based on the forecasting of \cite{Gilman:2019vca}, will enable us to either rule out or detect half-mode masses above $10^7$ M$_\odot$.

Of course, there is no a priori reason for the dark sector to be simple, no reason for DM to be a single entity. Indeed, the totality of the cosmological DM could be composed of all of the well-motivated WDM candidates, sterile neutrinos, axions, gravitinos, etc.   Being composed of different kinds of particles, such a mixed DM could have different clustering properties.  So we are interested in to what extent we can constrain such a complicated DM scenario with the large set of future strong gravitational lenses we will observe with JWST. Specifically for this paper, we are investigating a MixDM case where 50\% of the DM is cold and 50\% is warm.

We first implement the MixDM model using the cosmological software \texttt{CLASS}~\citep{CLASS}, which we use to calculate the transfer function for our model. The cosmology for these calculations was taken to be the Planck 2018 best-fit cosmology~\citep{Planck:2018vyg}. We take the best-fit density of CDM and then take half of that density and assign it to a WDM component.  We then calculate a halo mass function resulting from this transfer function using \texttt{galacticus}~\citep{galacticus} with the Sheth-Tormen~\citep{Sheth:1999mn} halo mass function.

Since the software we use to populate our lens models with DM halos, \texttt{pyHalo}, which renders full mass distributions for substructure lensing simulations with the open source gravitational lensing software package lenstronomy~\citep{lenstronomy}, requires a parameterized form of the halo mass function, we fit the output halo mass function from \texttt{CLASS}~\citep{CLASS} and \texttt{galacticus}~\citep{galacticus} as a suppression relative to the CDM mass function:
\begin{equation}
     \frac{dN/dM _{\rm MixDM}}{dN/dM _{\rm CDM}}(M) = (f + (1-f)(1 + a(M_\mathrm{hm}/M)^b)^{c/2})^2
\end{equation}
where, $a$, $b$, and $c$ are parameters describing the WDM suppression relative to CDM and we fix them to $a=0.5$, $b=0.8$, and $c=-3.0$. Also, $f$ is the fraction of CDM, which we set to 0.5 in our analyses. 

Intuitively, the $f^2$ in the mass function comes from the fact that there is an $f^2$ factor in the suppression in the power spectrum, since the power spectrum is a two-point statistic. Consider,
\begin{multline*}
\rho_{\rm DM} = f \rho_{\rm CDM} + (1-f) \rho_{\rm WDM} \\
\langle \rho_{\rm DM} \ \rho_{\rm DM}\rangle = f^2 \langle \rho_{\rm CDM} \ \rho_{\rm CDM}\rangle\\  + f(1-f) \langle \rho_{\rm CDM} \ \rho_{\rm WDM}\rangle + (1-f)^2\langle \rho_{\rm WDM} \ \rho_{\rm WDM}\rangle 
\end{multline*}
At sufficiently small scales, only the $f^2 \langle \rho_{\rm CDM} \ \rho_{\rm CDM} \rangle$ term remains.  Ultimately, this intuition is verified by our Galacticus calculations.

Further, the concentration-mass relation for MixDM is set to be identical to the WDM one which we use from \citet{Bose:2015mga}. This is a conservative choice when trying to differentiate MixDM and WDM. One would need to simulate a MixDM universe in order to robustly calculate the concentration-mass relation for this MixDM model.

In Fig.~\ref{fig:mixed_hmf}, we plot a MixDM case, WDM case, and CDM case assuming the Planck 2018 best-fit cosmology~\citep{Planck:2018vyg}.
As we see in Fig.~\ref{fig:mixed_hmf}, the MixDM case is suppressed relative to CDM, but there is no turnover and the lowest-mass halos are still abundant compared to the WDM case.  Since the strong lensing signal receives a contribution from a range of dark matter halo masses, the signal is less sensitive to the specific features of any model's halo mass function, but instead is sensitive to the total amount of suppression (see~Sec.~\ref{sec:SLDetails}). So we are investigating to what extent would WDM or MixDM scenarios be confused for one another.

\section{Strong Lensing Details}\label{sec:SLDetails}

Strong gravitational lenses are an exciting probe of DM physics as they can probe the mass function and structure of halos at cosmological distances regardless of whether they host galaxies. In a strong lens system, the positions and magnifications of the multiple images depend on the first and second derivatives of the gravitational potential, respectively.  The image positions typically offer a robust constraint on the model of the main deflector~\citep{Gilman:2016uit,Gilman:2017voy}.  The second derivative of the lens' gravitational potential is greatly altered by the presence of low-mass halos and thus the fluxes of the images are sensitive to them. The magnification field can be calculated at any point on the sky. Since the sources have finite sizes, the actual magnification of the images is integrated over a finite region of the sky.  Typically, smaller source sizes are sensitive to smaller halo masses since the effect of perturbations is integrated over the size of the source in the image plane.

As mentioned previously, we employ the flux ratio anomaly method in quadruply lensed images. Since the model of the main deflector gives a range of predictions for the fluxes of the four images, any additional DM halos that are satellites of the main deflector (subhalos) or halos along the line of sight would perturb the predictions of the four images' fluxes.  So this anomalous flux is the statistical signal that would give information about the halo mass function, the DM's transfer function, and ultimately about the particle identity of DM.

Narrow line emission from quasars has been the main source of data in the past~\citep{Nierenberg:2019pdj} and this will continue to improve as larger datasets at higher resolution and precision are acquired with improved Adaptive Optics and instruments such as KAPA \cite{KAPA} and LIGER \cite{LIGER} currently under development at Keck. Furthermore, the launch of JWST has opened up the new exciting possibility to use flux ratios measured in the mid-IR, where the source is typically smaller than the narrow line emission and thus more sensitive to low-mass perturbations.
The cold torus regions of quasars, detectable by JWST-MIRI, typically have source sizes in the range 1--10 pc, which is small enough to be sensitive to individual $10^7$ M$_\odot$ halos.  Thus, when we make our mock lensed images, we take a source size of 5 pc. 

Even in CDM, the processes by which subhalos and the satellite galaxies that inhabit them infall into and evolve within the host dark matter halo are very dynamical and non-linear and thus uncertain. Currently, we marginalize over this uncertainty with a parameter describing the normalization of the subhalo mass function, $\Sigma_\mathrm{sub}$, which is defined, following \cite{Gilman:2019nap}, such that,
\begin{equation}
\frac{d^2 N} {dM dA} = \frac{\Sigma_\mathrm{sub}}{M_0}\left(\frac{M}{M_0}\right)^\alpha \mathcal{F}(M_\mathrm{halo},z),
\end{equation}
where $\alpha$ is the slope of the subhalo mass function, the pivot halo mass is fixed to $M_0 = 10^8 M_\odot$, and $\mathcal{F}(M_\mathrm{halo},z)$ is a function that encapsulates the dependence of the subhalo mass function on host halo mass and redshift and is calibrated on results from \texttt{galacticus}. 
This parameter introduces large degeneracies in the inference of dark matter properties.

We model these strong lens systems with~\texttt{lenstronomy}~\cite{Birrer:2018, Birrer:2021wjl}.  Using the halo mass functions calculated with \texttt{galacticus}, we implement the MixDM model in \texttt{pyHalo}.

We generate 40 mock lenses from the MixDM model using a variety of main deflector configurations with different realizations of line-of-sight haloes. Subhalo populations were drawn from the MixDM model using parameters $M_\mathrm{hm}=10^{8.5} M_{\odot}$ and $\Sigma_\mathrm{sub}=0.05$ kpc$^{-2}$.

\section{Characterizing the MixDM Signal}\label{sec:Statistics}

\begin{figure*}
    \centering
    \includegraphics[width=\columnwidth]{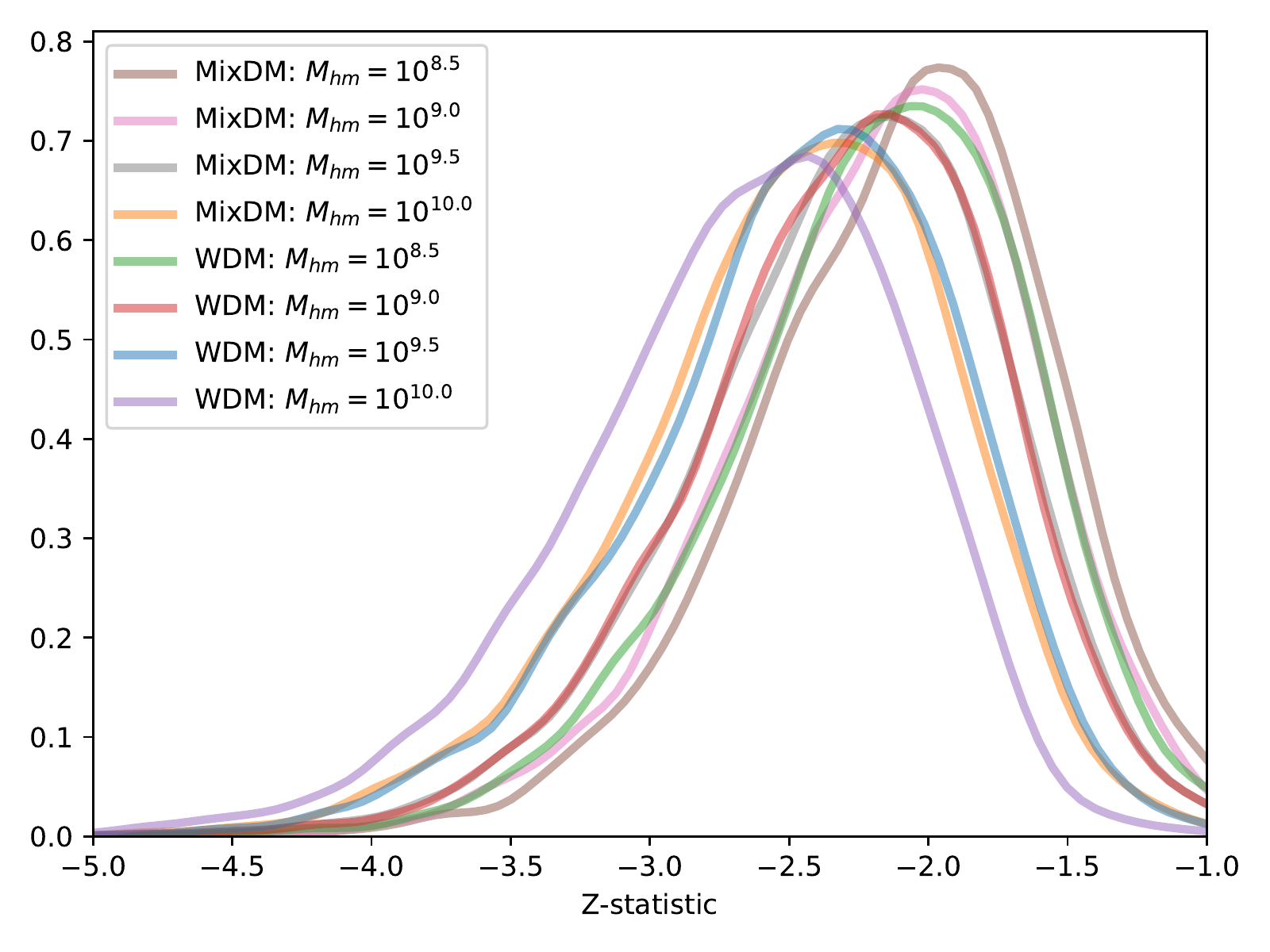}
    \includegraphics[width=\columnwidth]{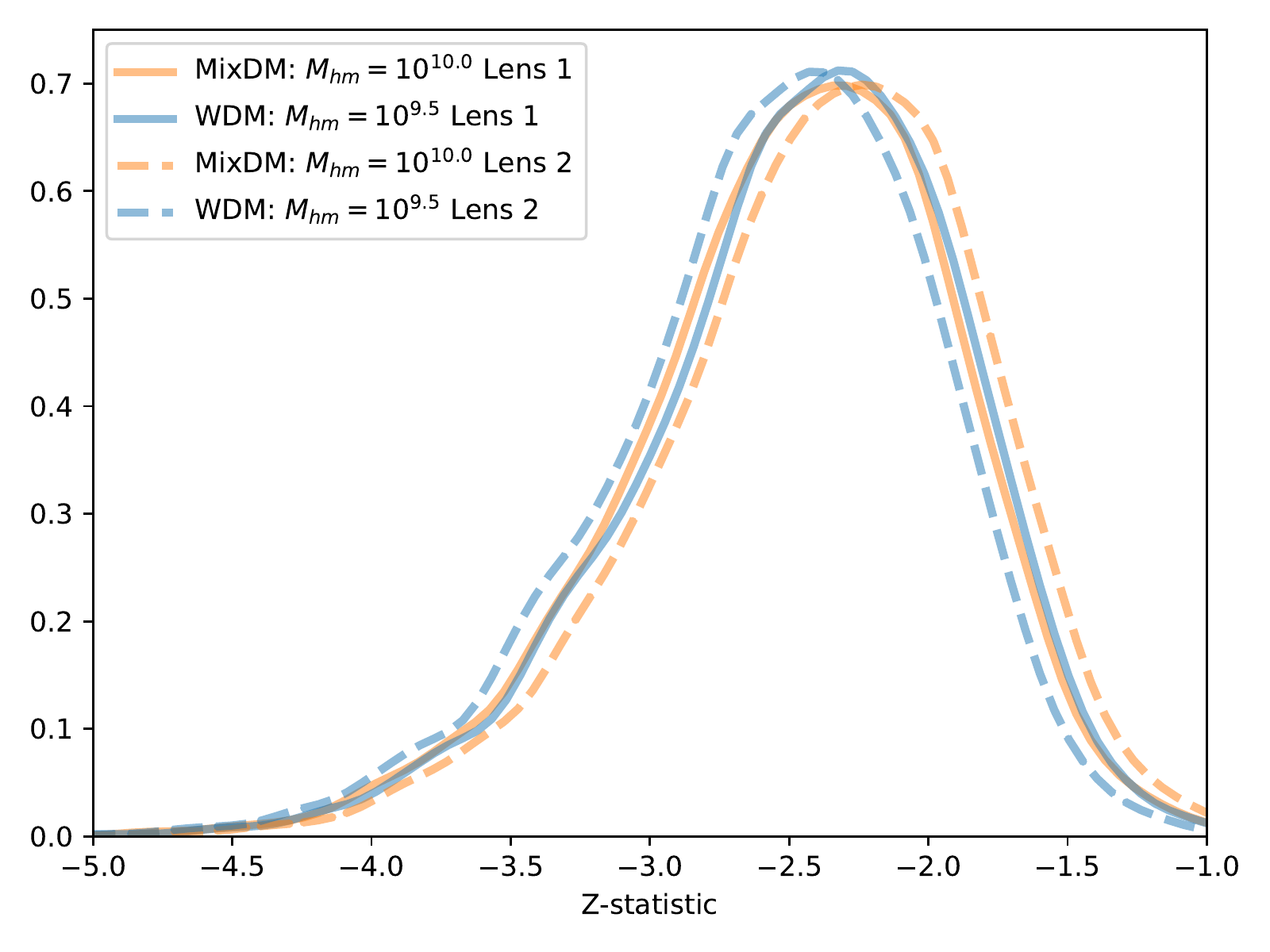}
    \caption{On the left, we show likelihoods of the Z-statistic for various half-mode masses for the WDM and MixDM cases for our first mock lens. Some of these distributions are identical, which would mean, no matter how many observations drawn from the distribution, they should be indistinguishable. On the right, we show likelihoods for $M_{\rm hm} = 10^{10} \mathrm{M}_\odot$ for the MixDM model and $M_{\rm hm} = 10^{9.5} \mathrm{M}_\odot$ for the WDM model, which overlap for the first lens, and likelihoods for the same parameters and models for the second lens, which do not over lap. Thus if we combine information from different lenses with different lens configurations will be able to break these degeneracies.}
    \label{fig:ex_dists}
\end{figure*}

\begin{figure}
    \centering
    \includegraphics[width=0.49\columnwidth]{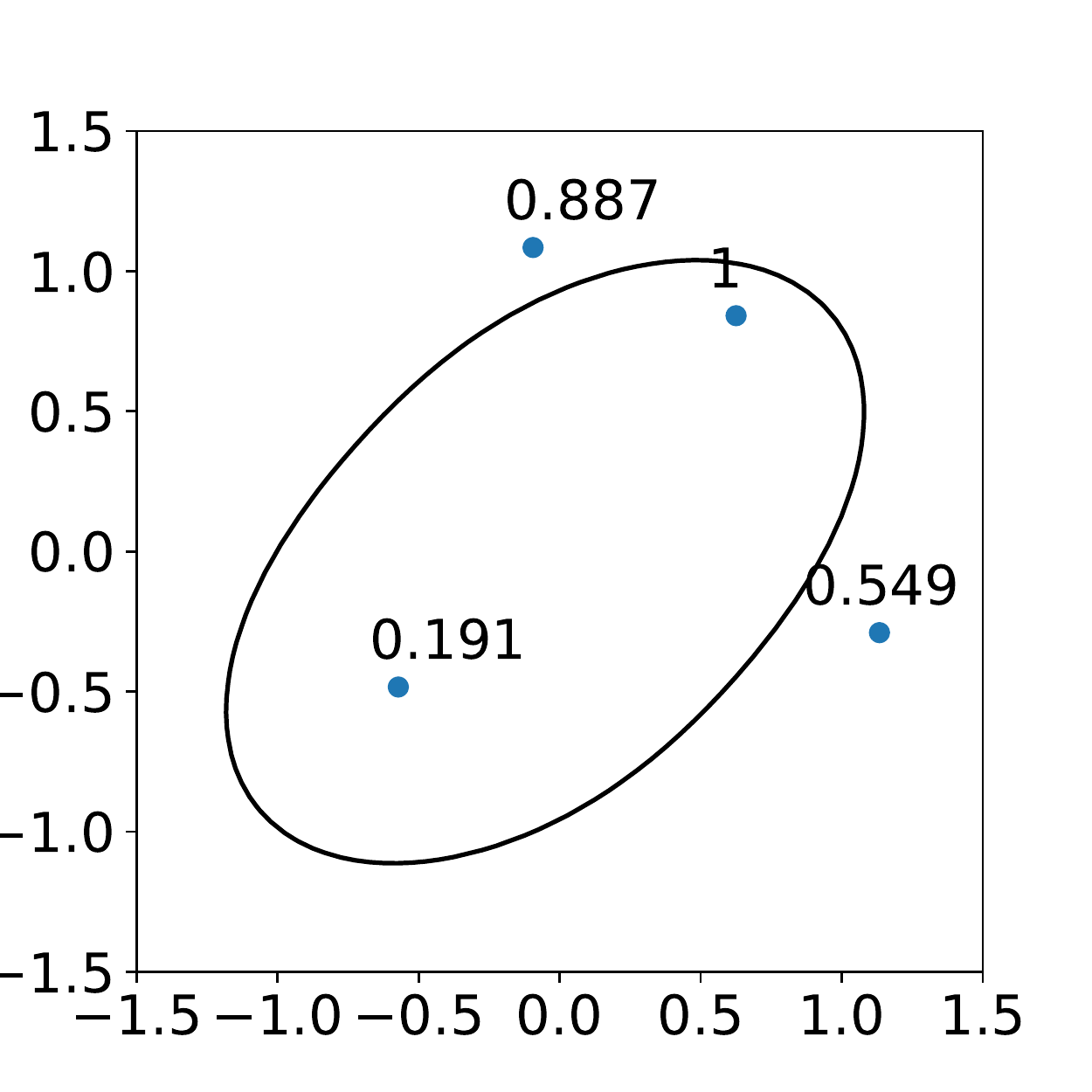}
    \includegraphics[width=0.49\columnwidth]{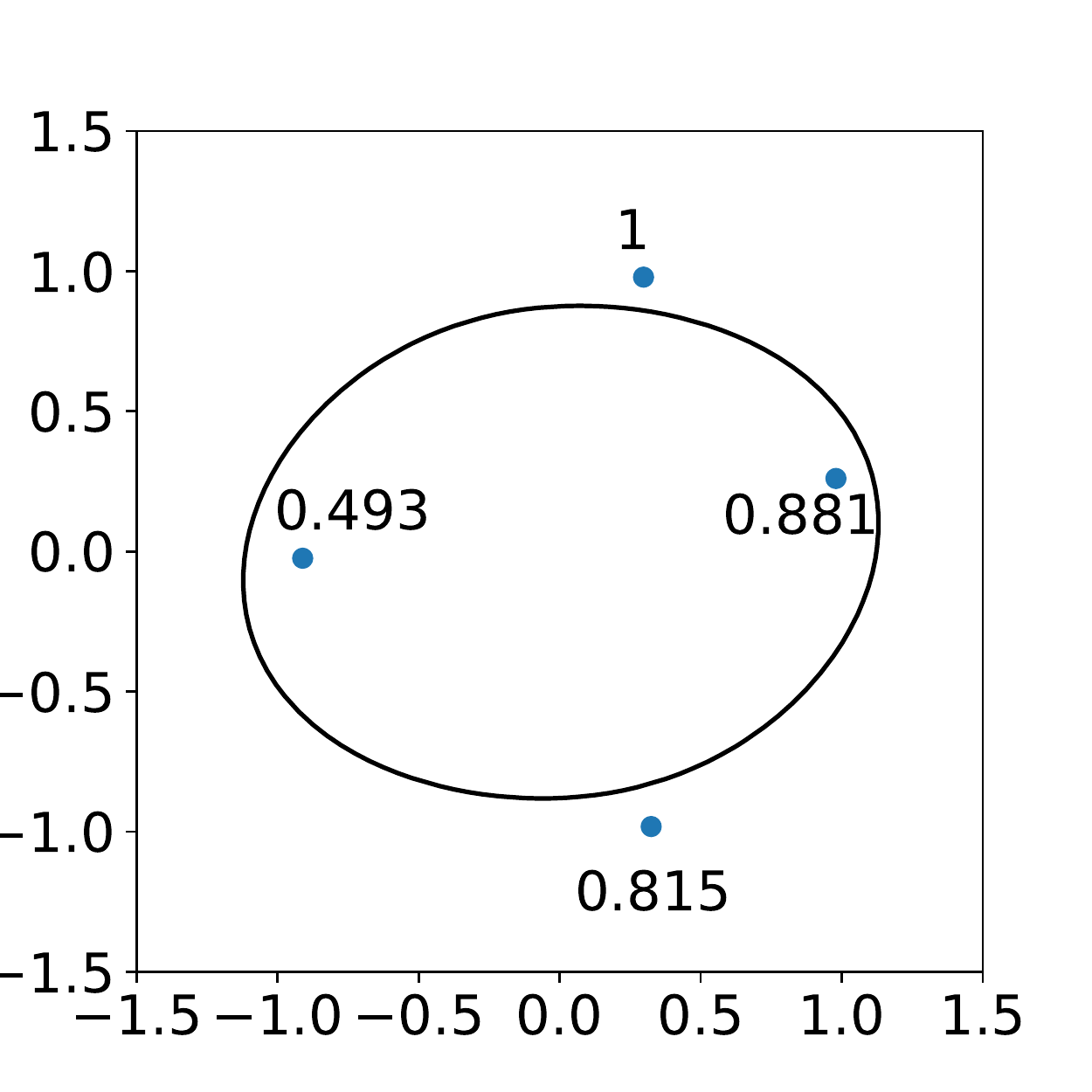}
    \caption{Configurations of the two lenses (lens 1 left, lens 2 right) in question and labelled by the flux ratios for each image. The critical curves are also displayed.}
    \label{fig:lensconf}
\end{figure}

We perform an initial test to determine whether the signals from the three DM models can be distinguished statistically.
Specifying the parameters of a DM model's halo mass function does not specify a single set of flux ratios, but instead a distribution of flux ratios because the way in which DM halos are configured around the lens and along the line-of-sight is not explicitly parametrized by the model.
Thus, inferring the properties of the halo mass function can only be done in a statistical manner,
as we are not directly counting the number of DM halos that give rise to the flux ratios,
it is prudent to calculate what a MixDM signal, compared to WDM, might look like. The signal of the difference between WDM and MixDM would be a different distribution in the predictions of the flux ratios.

In Fig.~\ref{fig:ex_ratios}, we see an example 2-dimensional histogram of flux ratios for CDM, WDM and MixDM.  The 2 dimensions in these figures are the flux ratios of the different images of a quadruply lensed system.  This distribution is over the different realizations of DM subhalos and line-of-sight halos. 
The histograms in Fig.~\ref{fig:ex_ratios} represent the distribution over possible observed flux ratios for CDM, WDM, and MixDM. 
The statistical signal that has the potential to differentiate the models is the scatter in the models' predictions, rather than the mean. 
The mean of the flux ratio anomaly method for all three models is similar, however, the distributions differ.
CDM has the largest variance in the predictions of the flux ratios, followed by MixDM and then WDM, which follows simply from the fact CDM predicts the largest number of DM subhalos, followed by MixDM and then WDM.

Further, we compress the full 3-D information of flux ratios into a summary statistic that characterizes how different the observed flux ratios are from the flux ratios predicted by the smooth main lens.
In particular, we compute these likelihoods by simulating the Z-statistic for a sampling of the different realizations of the distribution of line-of-sight and subhalos.

The Z-statistic is defined as follows:
\begin{equation}
    Z(f_i) = \sum_i (f_i - f_{\rm ref,i})^2 ,
\end{equation}
where $f_i$ are the flux ratios for the full model with the additional DM line-of-sight halos and subhalos, and $f_{\rm ref,i}$ is the reference image flux ratios that correspond to the predicted flux ratios from the macromodel of the main deflector lens when there is no additional DM line-of-sight halos or subhalos.  The sum is over each of the images.
We assume a fixed model of the main deflector and only allow for the realization of the DM substructure to vary. Similarly, for this initial test, we do not account for any statistical noise from the measurement of the fluxes of these quadruply lensed systems. Such noise would broaden these distributions in the same way between different models.
This is an idealistic case but a useful one for demonstrating how WDM and MixDM are statistically different.
We do this for different half-mode masses in the range for both the WDM and MixDM models.

In Fig.~\ref{fig:ex_dists}, we show the likelihood of the Z-statistic.  
We see that some of these likelihoods for WDM and MixDM are indistinguishable.  For instance, the likelihood for the MixDM model with $M_{\rm hm} = 10^{10} \ \mathrm{M}_\odot$ and WDM model with $M_{\rm hm} = 10^{9.5} \ \mathrm{M}_\odot$ lie on top of each other. 

This would imply that the two models, for these parameters, should be indistinguishable, no matter how many lenses are observed, and further, that even if DM were a mixture of warm and cold components, the corresponding suppression in the halo mass function could be interpreted as coming from a WDM model. However, these distributions were calculated for just a single lens configuration of the gravitational lens.  The range of halo masses that flux ratios are sensitive to will generally depend on the configuration of the lensed images in addition to the size of the source.
 
To this end, we additionally test if the degeneracy between WDM and MixDM is the same between different lens configurations.  Thus we perform the same calculation for a different mock lens configuration. In Fig.~\ref{fig:lensconf} we show the two example lenses we use to make this point.  We take the same WDM parameters ($M_\mathrm{hm}=10^{9.5} \mathrm{M}_\odot$) and MixDM parameters ($M_\mathrm{hm}=10^{10} \mathrm{M}_\odot$) that overlap in the first lens and calculate the distribution of their Z-statistics for the second lens.  We find that the likelihoods are different for our second lens configuration; the distributions shift in different directions for the second lens.
Thus, we find that a set of different lenses would at least weakly break any degeneracy between WDM and MixDM.

\section{Results}\label{sec:Results}

\begin{figure*}
    \centering
    \includegraphics[width=0.48\textwidth]{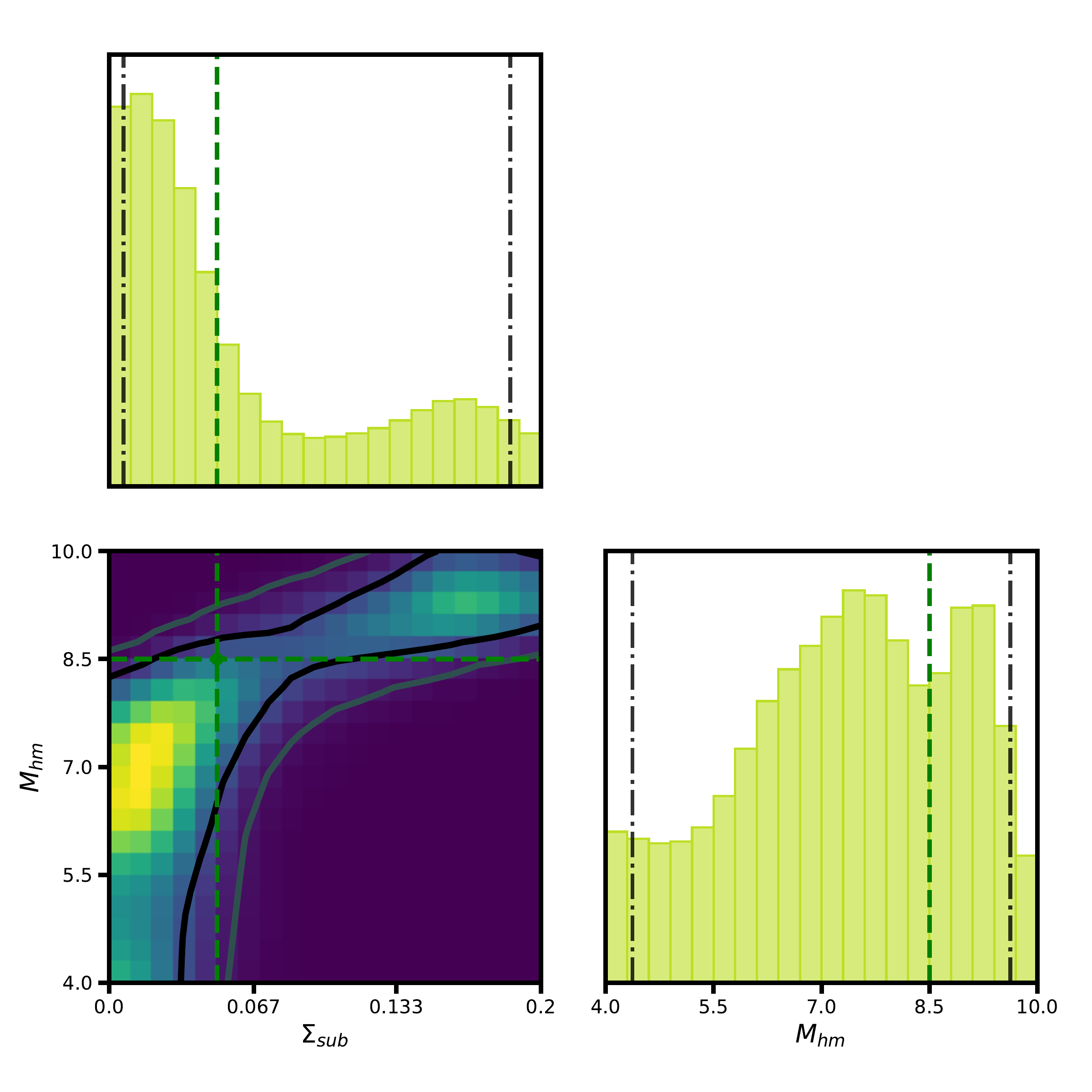}
    \includegraphics[width=0.48\textwidth]{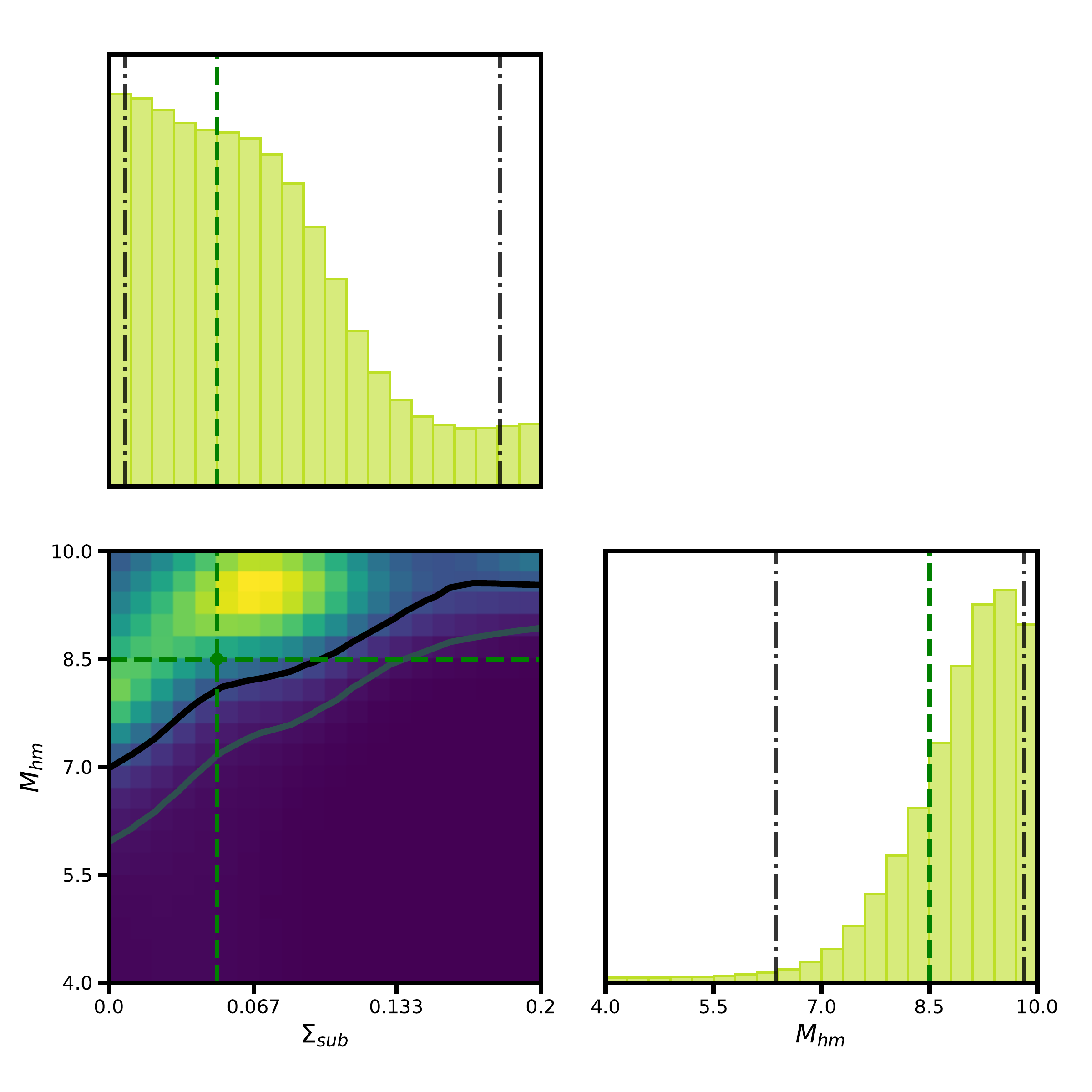}
    \caption{Posteriors for the half-mode mass and normalization of the subhalo mass function for 40 mock lenses for the WDM case (left) and for the MixDM case (right). The green line denotes the true parameters of the MixDM model that was used to generate the data.}
    \label{fig:posteriors}
\end{figure*}
 
Now that we have demonstrated 
that observing lenses with multiple, different configurations of main deflectors can break degeneracies between WDM and MixDM,
we want to be concrete about how many lenses are needed to differentiate between CDM, WDM, and MixDM.  

In order to do this calculation, we first create a set of flux ratio datasets from the MixDM model.  That is, we create a set of mock lens configurations and populate these lenses with additional DM halos generated statistically from the MixDM model.
Specifically, we choose $\log_{10} (M_{\rm hm}/\mathrm{M}_\odot) = 8.5$ and $\Sigma_{\rm sub} = 0.05$ kpc$^{-2}$ and $f=0.5$.  The flux ratios from these mock gravitational lens systems are our mock datasets.

The predictions of our strong lens modelling depends on not only the parameters of the lens and DM parameters, but also the specific configuration of DM subhalos and line-of-sight halos around the lens.  This configuration is not parametrized but is implemented as a stochastic process. Because the forward modelling involves a stochastic process, standard statistical inference techniques, such as MCMC or nested sampling, cannot be used.
Thus, we use approximate Bayesian computation (ABC) selection to approximate a posterior~\citep{Akeret:2015uha,Hahn:2016zwc, Birrer:2017}. This method was used to constraint DM models including WDM, SIDM in previous works including~\cite{Gilman:2016uit,Gilman:2017voy,Gilman:2019bdm,Gilman:2019vca,Gilman:2021sdr,Gilman:2022ida}.
ABC works by calculating a summary statistic ($S$) for how well the forward modeled flux ratios ($f_{i}$) match the data ($f_{\rm{data},i}$). Specifically, we choose an unweighted $\chi^2$ statistic
\begin{equation}
    S^2 = \sum_i(f_{i} - f_{\rm{data},i})^2 .
\end{equation}
We use this $S$-statistic for inferring strong lensing parameters, rather than the $Z$-statistic, to follow previous studies such as ~\cite{Gilman:2019vca,Gilman:2019bdm}, where it has been extensively tested on simulated data and shown to robustly recover input model parameters.  The ABC method then selects samples on the condition that the S-statistic is less than some threshold.  We choose $S<0.05$, which roughly corresponds to a $<3\%$ precision on individual flux ratios. We checked convergence by generating samples until the posterior stopped changing with increasing number of accepted samples.  We checked that the threshold doesn’t shift the posterior, just the contours become less noisy. We sample the prior uniformly in the range $\log_{10} ( M_{\rm hm} / \mathrm{M}_\odot) \in \{4,10\}$ and $\Sigma_{\rm sub} \in \{0.0, 0.2 \}$ kpc$^{-2}$. The S-statistic for each of these prior samples are calculated and if the S-statistic for a given sample is less than our threshold criteria of $S<0.05$, it is selected.  This set of selected samples parameters compose the set of parameters that can provide a reasonable fit to the data and approximate parameter samples drawn from the posterior. In order to combine these approximate posteriors for each lens into a joint posterior, we then calculate a kernel density estimate for the set of selected samples.  We calculate a joint posterior for the different lenses by simply multiplying the individual kernel density estimates.

Fig.~\ref{fig:posteriors} shows the results of these calculations, where we show the 1- and 2-$\sigma$ contours of the joint posteriors for our 40 mock lenses. The simple rule of thumb to interpret these posterior plots is that the bottom right corner, with parameters $\Sigma_\mathrm{sub}=0.2$ kpc$^{-2}$ and $\log_{10}M_\mathrm{hm}=4.0$ is the region with the most amount of structure, and the opposite upper left corner, with $\Sigma_\mathrm{sub}=0.0$ kpc$^{-2}$ and $\log_{10}M_\mathrm{hm}=10.0$ is the region with the least amount of structure. The true value for the parameters, as indicated by the green lines, was a case with an intermediate amount of structure, from the MixDM case.  The left panel shows the results for the WDM case and the right panel shows the case for the MixDM case.
For a fixed fraction of CDM ($f=0.5$), the MixDM model is necessarily less flexible than the WDM and predicts a narrower range of observables over its parameter space. For instance, in the MixDM case, the suppression in structure relative to CDM comes to a minimum value of the square of the CDM fraction ($f^2$) for small halo masses. This difference in the range of predictions for the two models is the explanation for the differences in the posteriors of the two models and what parameters they do and do not rule out. To explain, for the bottom right corner of parameter space, both models should have an large amount of structure that is observably indistinguishable, but the opposite, upper left corner for MixDM would predict categorically more structure than the WDM case. Since the true parameters correspond to roughly an intermediate amount of structure and the least amount of structure the MixDM can predict is an intermediate amount, the MixDM case cannot rule out the upper left corner.  In the WDM case, the upper left corner corresponds to very little amount of structure which the data can rule out.  Thus the contours are less well constrained in the MixDM case than in the WDM case.

\subsection{Bayesian Evidences}
\begin{figure}
    \centering
    \includegraphics[width=\columnwidth]{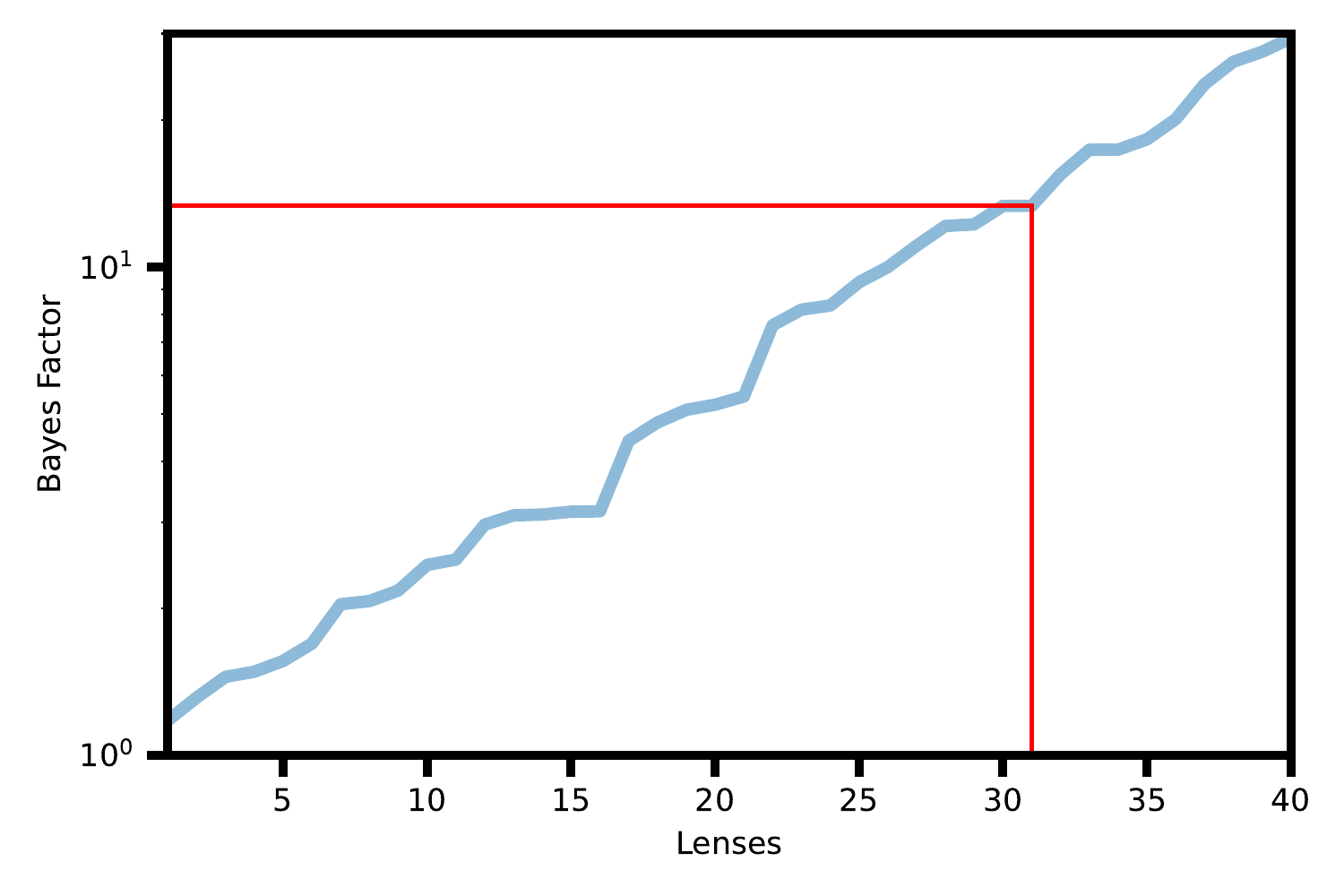}
    \caption{Cumulative Bayes factor between the MixDM and WDM models as a function of number of lenses. This surpasses the Jeffreys scale threshold for a ``strong'' preference at 25 lenses, and at 40 lenses, prefers the MixDM model by a factor of 30:1. The red line corresponds to the forecasted statistical preference for the 31 lenses we will observe with upcoming JWST observations.}
    \label{fig:bfs}
\end{figure}

The Bayesian evidence can also be approximated with the ABC method by simply calculating the frequency of a model satisfying the ABC criteria. Thus we can calculate a Bayes factor for a single lens by calculating how more frequently will samples from one model satisfy the ABC criteria than another. We then calculate the Bayes factor for the set of lenses by multiplying the Bayes factors for the individual lenses.  The result of this calculation is shown in Fig.~\ref{fig:bfs} where we can see that it takes around 35 lenses to achieve a 20:1 preference for MixDM over WDM for our mock MixDM and at 40 lenses, the preference is at a level of 30:1.  This relationship is linear in the $\log$ of the Bayes factor and so can be approximately extrapolated to a large number of lenses as would be expected from next generation surveys.

\subsection{Varying $f$}
\begin{figure}
    \centering
    \includegraphics[width=\columnwidth]{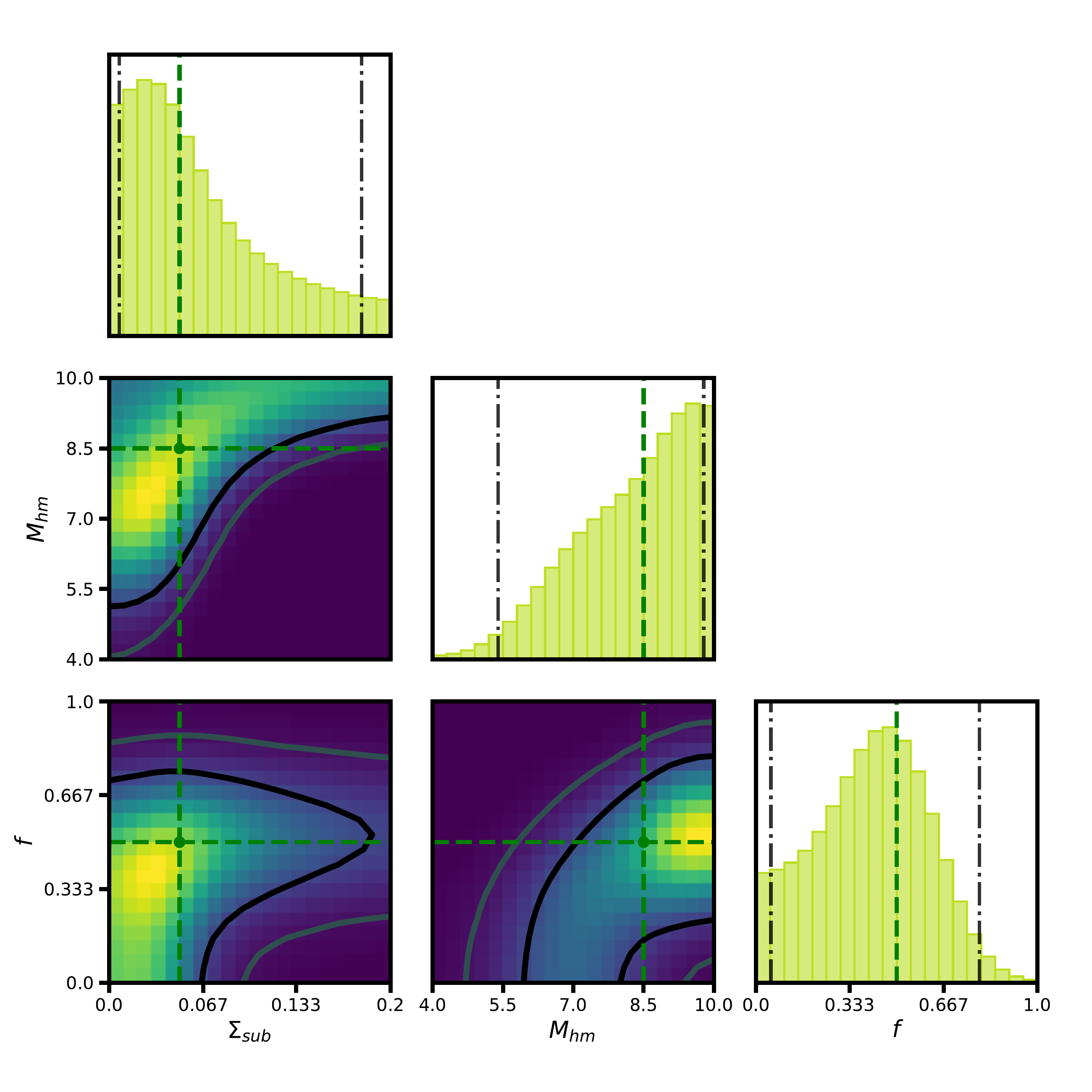}
    \caption{Posterior for the case where the CDM fraction $f$ is varied alongside the half-mode mass $M_{\rm hm}$ and the normalization of the subhalo mass function $\Sigma_{\rm sub}$.}
    \label{fig:varf}
\end{figure}

These previous results for the MixDM model were calculated with a fixed fraction of CDM ($f=0.5$) and now in Fig.~\ref{fig:varf} we calculate the posteriors for the full model case where where the fraction of CDM, $f$, is allowed to vary and the true value is $f=0.5$. Examining the marginalized posterior for $f$, we can see that the CDM regime $f=1$ is more easily ruled out than the WDM regime $f=0$.  This is a result of the only weakly broken degeneracies between CDM fraction, $f$, and the half-mode mass $M_\mathrm{hm}$.

\section{Conclusions}\label{sec:Conclusions}

In this paper, we have forecasted the potential for flux ratio anomalies in strong gravitational lensing to be sensitive to forms of suppression of the halo mass function beyond WDM.   To concretely demonstrate this point, we calculate the lensing signal from a MixDM model.

Since the anomalous flux method for detecting DM from strong lenses is sensitive to not just one mass scale of the halo mass function, but averaged over a range, it will be difficult to distinguish between different scenarios in which the halo mass function would be suppressed by different DM physics. Specifically, with only one lens configuration, a MixDM model could be confused for a WDM with a different half-mode mass. With a dataset composed of multiple different lens configurations, this degeneracy can be broken.  

One important caveat is that, at present, we have implemented the concentrations for MixDM halos using the WDM concentration-mass relation.  Since the MixDM concentration-mass relation is expected to be different, this can be a useful tool for differentiating the WDM and MixDM models from each other using strong lens systems.  Robustly calculating what the MixDM concentration-mass relation should be is left for future work.

Further, we find that, with 40 lenses, we can use the ABC method to distinguish between a MixDM model and a WDM with a confidence of 30:1 for the case when $f$ is fixed to 0.5. Further, with an upcoming JWST program (GO-2046) which will observe 31 lenses, a Bayes factor of 13:1 can be achieved, and thus, a MixDM model that is maximally different than CDM or WDM should be detectable, and more generally, such strong gravitational lens systems would have sensitivities to suppression in the halo mass function beyond just the WDM paradigm.  This prospect is exciting since, given that we can constrain these rather complex MixDM models, we might also be able to constrain other composite DM models or constrain features in the small-scale matter distribution.

\section*{Acknowledgements}
We would like to acknowledge all of the academic workers in the University of California system who have been striking to improve working and living conditions. 
This research was conducted using [MERCED cluster (NSF-MRI, \#1429783) / Pinnacles (NSF MRI, \#2019144) / Science DMZ (NSF-CC* \#1659210)] at the Cyberinfrastructure and Research Technologies (CIRT) at University of California, Merced.
TT and AN acknowledges support by NSF through grants AST-2205100 "Collaborative Research: Measurng the physical properties of dark matter with strong gravitational lensing", AST-1836016 "Astrophysics enabled by Keck All Sky Precision Adaptive Optics" and by the Gordon and Betty Moore Foundation through grant.
Support for program \#2046 was provided by NASA through a grant from the Space Telescope Institute, which is operated by the Association of Universities for Research in Astronomy, Inc., under NASA contract NAS 5-03127.

%%%%%%%%%%%%%%%%%%%%%%%%%%%%%%%%%%%%%%%%%%%%%%%%%%
\section*{Data Availability}
The data used in this work are mocks generated by the authors with the software packages \texttt{pyHalo} and \texttt{lenstronomy}, as described in the text.



\bibliographystyle{mnras}
\bibliography{example} % if your bibtex file is called example.bib

\begin{thebibliography}{}
\makeatletter
\relax
\def\mn@urlcharsother{\let\do\@makeother \do\$\do\&\do\#\do\^\do\_\do\%\do\~}
\def\mn@doi{\begingroup\mn@urlcharsother \@ifnextchar [ {\mn@doi@}
  {\mn@doi@[]}}
\def\mn@doi@[#1]#2{\def\@tempa{#1}\ifx\@tempa\@empty \href
  {http://dx.doi.org/#2} {doi:#2}\else \href {http://dx.doi.org/#2} {#1}\fi
  \endgroup}
\def\mn@eprint#1#2{\mn@eprint@#1:#2::\@nil}
\def\mn@eprint@arXiv#1{\href {http://arxiv.org/abs/#1} {{\tt arXiv:#1}}}
\def\mn@eprint@dblp#1{\href {http://dblp.uni-trier.de/rec/bibtex/#1.xml}
  {dblp:#1}}
\def\mn@eprint@#1:#2:#3:#4\@nil{\def\@tempa {#1}\def\@tempb {#2}\def\@tempc
  {#3}\ifx \@tempc \@empty \let \@tempc \@tempb \let \@tempb \@tempa \fi \ifx
  \@tempb \@empty \def\@tempb {arXiv}\fi \@ifundefined
  {mn@eprint@\@tempb}{\@tempb:\@tempc}{\expandafter \expandafter \csname
  mn@eprint@\@tempb\endcsname \expandafter{\@tempc}}}

\bibitem[\protect\citeauthoryear{Abazajian}{Abazajian}{2017}]{Abazajian:2017tcc}
Abazajian K.~N.,  2017, \mn@doi [Phys. Rept.] {10.1016/j.physrep.2017.10.003},
  711-712, 1

\bibitem[\protect\citeauthoryear{Abazajian \& Kusenko}{Abazajian \&
  Kusenko}{2019}]{Abazajian:2019ejt}
Abazajian K.~N.,  Kusenko A.,  2019, \mn@doi [Phys. Rev. D]
  {10.1103/PhysRevD.100.103513}, 100, 103513

\bibitem[\protect\citeauthoryear{Abazajian, Fuller  \& Patel}{Abazajian
  et~al.}{2001}]{Abazajian:2001nj}
Abazajian K.,  Fuller G.~M.,   Patel M.,  2001, \mn@doi [Phys. Rev. D]
  {10.1103/PhysRevD.64.023501}, 64, 023501

\bibitem[\protect\citeauthoryear{Abazajian, Horiuchi, Kaplinghat, Keeley  \&
  Macias}{Abazajian et~al.}{2020}]{Abazajian:2020tww}
Abazajian K.~N.,  Horiuchi S.,  Kaplinghat M.,  Keeley R.~E.,   Macias O.,
  2020, \mn@doi [Phys. Rev. D] {10.1103/PhysRevD.102.043012}, 102, 043012

\bibitem[\protect\citeauthoryear{Aghanim et~al.}{Aghanim
  et~al.}{2020}]{Planck:2018vyg}
Aghanim N.,  et~al., 2020, \mn@doi [Astron. Astrophys.]
  {10.1051/0004-6361/201833910}, 641, A6

\bibitem[\protect\citeauthoryear{Akeret, Refregier, Amara, Seehars  \&
  Hasner}{Akeret et~al.}{2015}]{Akeret:2015uha}
Akeret J.,  Refregier A.,  Amara A.,  Seehars S.,   Hasner C.,  2015, \mn@doi
  [JCAP] {10.1088/1475-7516/2015/08/043}, 08, 043

\bibitem[\protect\citeauthoryear{Amara, Metcalf, Cox  \& P.}{Amara
  et~al.}{2006}]{Amara:2004dr}
Amara A.,  Metcalf R.~B.,  Cox T.~J.,   P. O.~J.,  2006, \mn@doi [Mon. Not.
  Roy. Astron. Soc.] {10.1111/j.1365-2966.2006.10053.x}, 367, 1367

\bibitem[\protect\citeauthoryear{{Anderhalden}, {Diemand}, {Bertone},
  {Macci{\`o}}  \& {Schneider}}{{Anderhalden}
  et~al.}{2012}]{2012JCAP...10..047A}
{Anderhalden} D.,  {Diemand} J.,  {Bertone} G.,  {Macci{\`o}} A.~V.,
  {Schneider} A.,  2012, \mn@doi [\jcap] {10.1088/1475-7516/2012/10/047}, \href
  {https://ui.adsabs.harvard.edu/abs/2012JCAP...10..047A} {2012, 047}

\bibitem[\protect\citeauthoryear{Andrade, Minor, Nierenberg  \&
  Kaplinghat}{Andrade et~al.}{2019}]{Andrade:2019wzn}
Andrade K.~E.,  Minor Q.,  Nierenberg A.,   Kaplinghat M.,  2019, \mn@doi [Mon.
  Not. Roy. Astron. Soc.] {10.1093/mnras/stz1360}, 487, 1905

\bibitem[\protect\citeauthoryear{Andrade, Fuson, Gad-Nasr, Kong, Minor, Roberts
   \& Kaplinghat}{Andrade et~al.}{2021}]{Andrade:2020lqq}
Andrade K.~E.,  Fuson J.,  Gad-Nasr S.,  Kong D.,  Minor Q.,  Roberts M.~G.,
  Kaplinghat M.,  2021, \mn@doi [Mon. Not. Roy. Astron. Soc.]
  {10.1093/mnras/stab3241}, 510, 54

\bibitem[\protect\citeauthoryear{Angulo, Hahn, Ludlow  \& Bonoli}{Angulo
  et~al.}{2017}]{Angulo:2016qof}
Angulo R.~E.,  Hahn O.,  Ludlow A.,   Bonoli S.,  2017, \mn@doi [Mon. Not. Roy.
  Astron. Soc.] {10.1093/mnras/stx1658}, 471, 4687

\bibitem[\protect\citeauthoryear{{Benson}}{{Benson}}{2012}]{galacticus}
{Benson} A.~J.,  2012, \na, \href
  {https://ui.adsabs.harvard.edu/abs/2012NewA...17..175B} {17, 175}

\bibitem[\protect\citeauthoryear{{Birrer} \& {Amara}}{{Birrer} \&
  {Amara}}{2018}]{Birrer:2018}
{Birrer} S.,  {Amara} A.,  2018, \mn@doi [Physics of the Dark Universe]
  {10.1016/j.dark.2018.11.002}, \href
  {https://ui.adsabs.harvard.edu/abs/2018PDU....22..189B} {22, 189}

\bibitem[\protect\citeauthoryear{{Birrer}, {Amara}  \& {Refregier}}{{Birrer}
  et~al.}{2017}]{Birrer:2017}
{Birrer} S.,  {Amara} A.,   {Refregier} A.,  2017, \mn@doi [\jcap]
  {10.1088/1475-7516/2017/05/037}, \href
  {https://ui.adsabs.harvard.edu/abs/2017JCAP...05..037B} {2017, 037}

\bibitem[\protect\citeauthoryear{{Birrer} et~al.}{{Birrer}
  et~al.}{2018}]{lenstronomy}
{Birrer} S.,  et~al., 2018, PotDU, \href
  {https://ui.adsabs.harvard.edu/abs/2018PDU....22..189B} {22, 189}

\bibitem[\protect\citeauthoryear{Birrer et~al.}{Birrer
  et~al.}{2021}]{Birrer:2021wjl}
Birrer S.,  et~al., 2021, \mn@doi [J. Open Source Softw.]
  {10.21105/joss.03283}, 6, 3283

\bibitem[\protect\citeauthoryear{{Bond} et~al.}{{Bond}
  et~al.}{1983}]{1983ApJ...274..443B}
{Bond} J.~R.,  et~al., 1983, \apj, \href
  {https://ui.adsabs.harvard.edu/abs/1983ApJ...274..443B} {274, 443}

\bibitem[\protect\citeauthoryear{Bose, Hellwing, Frenk, Jenkins, Lovell, Helly
  \& Li}{Bose et~al.}{2016}]{Bose:2015mga}
Bose S.,  Hellwing W.~A.,  Frenk C.~S.,  Jenkins A.,  Lovell M.~R.,  Helly
  J.~C.,   Li B.,  2016, \mn@doi [Mon. Not. Roy. Astron. Soc.]
  {10.1093/mnras/stv2294}, 455, 318

\bibitem[\protect\citeauthoryear{Boyarsky, Lesgourgues, Ruchayskiy  \&
  Viel}{Boyarsky et~al.}{2009}]{Boyarsky:2008xj}
Boyarsky A.,  Lesgourgues J.,  Ruchayskiy O.,   Viel M.,  2009, \mn@doi [JCAP]
  {10.1088/1475-7516/2009/05/012}, 05, 012

\bibitem[\protect\citeauthoryear{Chen, Kravtsov  \& Keeton}{Chen
  et~al.}{2003}]{Chen:2003uu}
Chen J.,  Kravtsov A.~V.,   Keeton C.~R.,  2003, \mn@doi [Astrophys. J.]
  {10.1086/375639}, 592, 24

\bibitem[\protect\citeauthoryear{{Colombi} et~al.}{{Colombi}
  et~al.}{1996}]{1996ApJ...458....1C}
{Colombi} S.,  et~al., 1996, \apj, \href
  {https://ui.adsabs.harvard.edu/abs/1996ApJ...458....1C} {458, 1}

\bibitem[\protect\citeauthoryear{Dalal \& Kochanek}{Dalal \&
  Kochanek}{2002}]{Dalal:2001fq}
Dalal N.,  Kochanek C.~S.,  2002, \mn@doi [Astrophys. J.] {10.1086/340303},
  572, 25

\bibitem[\protect\citeauthoryear{Despali, Giocoli, Angulo, Tormen, Sheth, Baso
  \& Moscardini}{Despali et~al.}{2016}]{Despali:2015yla}
Despali G.,  Giocoli C.,  Angulo R.~E.,  Tormen G.,  Sheth R.~K.,  Baso G.,
  Moscardini L.,  2016, \mn@doi [Mon. Not. Roy. Astron. Soc.]
  {10.1093/mnras/stv2842}, 456, 2486

\bibitem[\protect\citeauthoryear{Dodelson \& Widrow}{Dodelson \&
  Widrow}{1994}]{Dodelson:1993je}
Dodelson S.,  Widrow L.~M.,  1994, \mn@doi [Phys. Rev. Lett.]
  {10.1103/PhysRevLett.72.17}, 72, 17

\bibitem[\protect\citeauthoryear{{Dodelson} et~al.}{{Dodelson}
  et~al.}{1994}]{1994PhRvL..72...17D}
{Dodelson} S.,  et~al., 1994, \prl, \href
  {https://ui.adsabs.harvard.edu/abs/1994PhRvL..72...17D} {72, 17}

\bibitem[\protect\citeauthoryear{Dutton \& Macci\`o}{Dutton \&
  Macci\`o}{2014}]{Dutton:2014xda}
Dutton A.~A.,  Macci\`o A.~V.,  2014, \mn@doi [Mon. Not. Roy. Astron. Soc.]
  {10.1093/mnras/stu742}, 441, 3359

\bibitem[\protect\citeauthoryear{{Enzi} et~al.,}{{Enzi}
  et~al.}{2021}]{2021MNRAS.506.5848E}
{Enzi} W.,  et~al., 2021, \mn@doi [\mnras] {10.1093/mnras/stab1960}, \href
  {https://ui.adsabs.harvard.edu/abs/2021MNRAS.506.5848E} {506, 5848}

\bibitem[\protect\citeauthoryear{Gaskins}{Gaskins}{2016}]{Gaskins:2016cha}
Gaskins J.~M.,  2016, \mn@doi [Contemp. Phys.] {10.1080/00107514.2016.1175160},
  57, 496

\bibitem[\protect\citeauthoryear{Gilman, Agnello, Treu, Keeton  \&
  Nierenberg}{Gilman et~al.}{2017}]{Gilman:2016uit}
Gilman D.,  Agnello A.,  Treu T.,  Keeton C.~R.,   Nierenberg A.~M.,  2017,
  \mn@doi [Mon. Not. Roy. Astron. Soc.] {10.1093/mnras/stx158}, 467, 3970

\bibitem[\protect\citeauthoryear{Gilman, Birrer, Treu, Keeton  \&
  Nierenberg}{Gilman et~al.}{2018}]{Gilman:2017voy}
Gilman D.,  Birrer S.,  Treu T.,  Keeton C.~R.,   Nierenberg A.,  2018, \mn@doi
  [Mon. Not. Roy. Astron. Soc.] {10.1093/mnras/sty2261}, 481, 819

\bibitem[\protect\citeauthoryear{Gilman, Birrer, Treu, Nierenberg  \&
  Benson}{Gilman et~al.}{2019}]{Gilman:2019vca}
Gilman D.,  Birrer S.,  Treu T.,  Nierenberg A.,   Benson A.,  2019, \mn@doi
  [Mon. Not. Roy. Astron. Soc.] {10.1093/mnras/stz1593}, 487, 5721

\bibitem[\protect\citeauthoryear{Gilman, Birrer, Nierenberg, Treu, Du  \&
  Benson}{Gilman et~al.}{2020a}]{Gilman:2019nap}
Gilman D.,  Birrer S.,  Nierenberg A.,  Treu T.,  Du X.,   Benson A.,  2020a,
  \mn@doi [Mon. Not. Roy. Astron. Soc.] {10.1093/mnras/stz3480}, 491, 6077

\bibitem[\protect\citeauthoryear{Gilman, Du, Benson, Birrer, Nierenberg  \&
  Treu}{Gilman et~al.}{2020b}]{Gilman:2019bdm}
Gilman D.,  Du X.,  Benson A.,  Birrer S.,  Nierenberg A.,   Treu T.,  2020b,
  \mn@doi [Mon. Not. Roy. Astron. Soc.] {10.1093/mnrasl/slz173}, 492, L12

\bibitem[\protect\citeauthoryear{Gilman, Bovy, Treu, Nierenberg, Birrer, Benson
   \& Sameie}{Gilman et~al.}{2021}]{Gilman:2021sdr}
Gilman D.,  Bovy J.,  Treu T.,  Nierenberg A.,  Birrer S.,  Benson A.,   Sameie
  O.,  2021, \mn@doi [Mon. Not. Roy. Astron. Soc.] {10.1093/mnras/stab2335},
  507, 2432

\bibitem[\protect\citeauthoryear{Gilman, Zhong  \& Bovy}{Gilman
  et~al.}{2022}]{Gilman:2022ida}
Gilman D.,  Zhong Y.-M.,   Bovy J.,  2022, arXiv e-prints

\bibitem[\protect\citeauthoryear{{Green} et~al.}{{Green}
  et~al.}{2004}]{2004MNRAS.353L..23G}
{Green} A.~M.,  et~al., 2004, \mnras, \href
  {https://ui.adsabs.harvard.edu/abs/2004MNRAS.353L..23G} {353, L23}

\bibitem[\protect\citeauthoryear{Hahn, Vakili, Walsh, Hearin, Hogg  \&
  Campbell}{Hahn et~al.}{2017}]{Hahn:2016zwc}
Hahn C.,  Vakili M.,  Walsh K.,  Hearin A.~P.,  Hogg D.~W.,   Campbell D.,
  2017, \mn@doi [Mon. Not. Roy. Astron. Soc.] {10.1093/mnras/stx894}, 469, 2791

\bibitem[\protect\citeauthoryear{{Hsueh} et~al.}{{Hsueh}
  et~al.}{2020}]{2020MNRAS.492.3047H}
{Hsueh} J.~W.,  et~al., 2020, \mnras, \href
  {https://ui.adsabs.harvard.edu/abs/2020MNRAS.492.3047H} {492, 3047}

\bibitem[\protect\citeauthoryear{Kahlhoefer}{Kahlhoefer}{2017}]{Kahlhoefer:2017dnp}
Kahlhoefer F.,  2017, \mn@doi [Int. J. Mod. Phys. A]
  {10.1142/S0217751X1730006X}, 32, 1730006

\bibitem[\protect\citeauthoryear{Kamada, Inoue  \& Takahashi}{Kamada
  et~al.}{2016}]{Kamada:2016vsc}
Kamada A.,  Inoue K.~T.,   Takahashi T.,  2016, \mn@doi [Phys. Rev. D]
  {10.1103/PhysRevD.94.023522}, 94, 023522

\bibitem[\protect\citeauthoryear{{Kim} \& {Peter}}{{Kim} \&
  {Peter}}{2021}]{2021arXiv210609050K}
{Kim} S.~Y.,  {Peter} A. H.~G.,  2021, arXiv e-prints, \href
  {https://ui.adsabs.harvard.edu/abs/2021arXiv210609050K} {p. arXiv:2106.09050}

\bibitem[\protect\citeauthoryear{Kim, Peter  \& Hargis}{Kim
  et~al.}{2018}]{Kim:2017iwr}
Kim S.~Y.,  Peter A. H.~G.,   Hargis J.~R.,  2018, \mn@doi [Phys. Rev. Lett.]
  {10.1103/PhysRevLett.121.211302}, 121, 211302

\bibitem[\protect\citeauthoryear{Kusenko}{Kusenko}{2009}]{Kusenko:2009up}
Kusenko A.,  2009, \mn@doi [Phys. Rept.] {10.1016/j.physrep.2009.07.004}, 481,
  1

\bibitem[\protect\citeauthoryear{{Laroche}, {Gilman}, {Li}, {Bovy}  \&
  {Du}}{{Laroche} et~al.}{2022}]{2022MNRAS.517.1867L}
{Laroche} A.,  {Gilman} D.,  {Li} X.,  {Bovy} J.,   {Du} X.,  2022, \mn@doi
  [\mnras] {10.1093/mnras/stac2677}, \href
  {https://ui.adsabs.harvard.edu/abs/2022MNRAS.517.1867L} {517, 1867}

\bibitem[\protect\citeauthoryear{{Lesgourgues}}{{Lesgourgues}}{2011}]{CLASS}
{Lesgourgues} J.,  2011, arXiv e-prints, \href
  {https://ui.adsabs.harvard.edu/abs/2011arXiv1104.2932L} {p. arXiv:1104.2932}

\bibitem[\protect\citeauthoryear{Mao \& Schneider}{Mao \&
  Schneider}{1998}]{Mao:1997ek}
Mao S.-d.,  Schneider P.,  1998, \mn@doi [Mon. Not. Roy. Astron. Soc.]
  {10.1046/j.1365-8711.1998.01319.x}, 295, 587

\bibitem[\protect\citeauthoryear{Metcalf}{Metcalf}{2005}]{Metcalf:2004eh}
Metcalf R.~B.,  2005, \mn@doi [Astrophys. J.] {10.1086/431574}, 629, 673

\bibitem[\protect\citeauthoryear{Metcalf \& Madau}{Metcalf \&
  Madau}{2001}]{Metcalf:2001ap}
Metcalf R.~B.,  Madau P.,  2001, \mn@doi [Astrophys. J.] {10.1086/323695}, 563,
  9

\bibitem[\protect\citeauthoryear{Minezaki, Chiba, Kashikawa, Inoue  \&
  Kataza}{Minezaki et~al.}{2009}]{Minezaki:2009ek}
Minezaki T.,  Chiba M.,  Kashikawa N.,  Inoue K.~T.,   Kataza H.,  2009,
  \mn@doi [Astrophys. J.] {10.1088/0004-637X/697/1/610}, 697, 610

\bibitem[\protect\citeauthoryear{Minor, Kaplinghat, Chan  \& Simon}{Minor
  et~al.}{2021a}]{Minor:2020bmp}
Minor Q.~E.,  Kaplinghat M.,  Chan T.~H.,   Simon E.,  2021a, \mn@doi [Mon.
  Not. Roy. Astron. Soc.] {10.1093/mnras/stab2209}, 507, 1202

\bibitem[\protect\citeauthoryear{Minor, Gad-Nasr, Kaplinghat  \& Vegetti}{Minor
  et~al.}{2021b}]{Minor:2020hic}
Minor Q.~E.,  Gad-Nasr S.,  Kaplinghat M.,   Vegetti S.,  2021b, \mn@doi [Mon.
  Not. Roy. Astron. Soc.] {10.1093/mnras/stab2247}, 507, 1662

\bibitem[\protect\citeauthoryear{Miranda \& Maccio}{Miranda \&
  Maccio}{2007}]{Miranda:2007rb}
Miranda M.,  Maccio A.~V.,  2007, \mn@doi [Mon. Not. Roy. Astron. Soc.]
  {10.1111/j.1365-2966.2007.12440.x}, 382, 1225

\bibitem[\protect\citeauthoryear{Moustakas \& Metcalf}{Moustakas \&
  Metcalf}{2003}]{Moustakas:2002iz}
Moustakas L.~A.,  Metcalf R.~B.,  2003, \mn@doi [Mon. Not. Roy. Astron. Soc.]
  {10.1046/j.1365-8711.2003.06055.x}, 339, 607

\bibitem[\protect\citeauthoryear{Nadler, Birrer, Gilman, Wechsler, Du, Benson,
  Nierenberg  \& Treu}{Nadler et~al.}{2021}]{Nadler:2021dft}
Nadler E.~O.,  Birrer S.,  Gilman D.,  Wechsler R.~H.,  Du X.,  Benson A.,
  Nierenberg A.~M.,   Treu T.,  2021, \mn@doi [Astrophys. J.]
  {10.3847/1538-4357/abf9a3}, 917, 7

\bibitem[\protect\citeauthoryear{{Niemeyer}}{{Niemeyer}}{2020}]{2020PrPNP.11303787N}
{Niemeyer} J.~C.,  2020, \mn@doi [Progress in Particle and Nuclear Physics]
  {10.1016/j.ppnp.2020.103787}, \href
  {https://ui.adsabs.harvard.edu/abs/2020PrPNP.11303787N} {113, 103787}

\bibitem[\protect\citeauthoryear{Nierenberg, Treu, Menci, Lu, Torrey  \&
  Vogelsberger}{Nierenberg et~al.}{2016}]{Nierenberg:2016nri}
Nierenberg A.~M.,  Treu T.,  Menci N.,  Lu Y.,  Torrey P.,   Vogelsberger M.,
  2016, \mn@doi [Mon. Not. Roy. Astron. Soc.] {10.1093/mnras/stw1860}, 462,
  4473

\bibitem[\protect\citeauthoryear{Nierenberg et~al.}{Nierenberg
  et~al.}{2020}]{Nierenberg:2019pdj}
Nierenberg A.~M.,  et~al., 2020, \mn@doi [Mon. Not. Roy. Astron. Soc.]
  {10.1093/mnras/stz3588}, 492, 5314

\bibitem[\protect\citeauthoryear{Parimbelli, Scelfo, Giri, Schneider,
  Archidiacono, Camera  \& Viel}{Parimbelli et~al.}{2021}]{Parimbelli:2021mtp}
Parimbelli G.,  Scelfo G.,  Giri S.~K.,  Schneider A.,  Archidiacono M.,
  Camera S.,   Viel M.,  2021, \mn@doi [JCAP] {10.1088/1475-7516/2021/12/044},
  12, 044

\bibitem[\protect\citeauthoryear{Patwardhan, Fuller, Kishimoto  \&
  Kusenko}{Patwardhan et~al.}{2015}]{Patwardhan:2015kga}
Patwardhan A.~V.,  Fuller G.~M.,  Kishimoto C.~T.,   Kusenko A.,  2015, \mn@doi
  [Phys. Rev. D] {10.1103/PhysRevD.92.103509}, 92, 103509

\bibitem[\protect\citeauthoryear{Robles, Bullock  \& Boylan-Kolchin}{Robles
  et~al.}{2019}]{Robles:2018fur}
Robles V.~H.,  Bullock J.~S.,   Boylan-Kolchin M.,  2019, \mn@doi [Mon. Not.
  Roy. Astron. Soc.] {10.1093/mnras/sty3190}, 483, 289

\bibitem[\protect\citeauthoryear{Schumann}{Schumann}{2019}]{Schumann:2019eaa}
Schumann M.,  2019, \mn@doi [J. Phys. G] {10.1088/1361-6471/ab2ea5}, 46, 103003

\bibitem[\protect\citeauthoryear{Sheth \& Tormen}{Sheth \&
  Tormen}{1999}]{Sheth:1999mn}
Sheth R.~K.,  Tormen G.,  1999, \mn@doi [Mon. Not. Roy. Astron. Soc.]
  {10.1046/j.1365-8711.1999.02692.x}, 308, 119

\bibitem[\protect\citeauthoryear{Shi \& Fuller}{Shi \&
  Fuller}{1999}]{Shi:1998km}
Shi X.-D.,  Fuller G.~M.,  1999, \mn@doi [Phys. Rev. Lett.]
  {10.1103/PhysRevLett.82.2832}, 82, 2832

\bibitem[\protect\citeauthoryear{{Shi} et~al.}{{Shi}
  et~al.}{1999}]{1999PhRvL..82.2832S}
{Shi} X.,  et~al., 1999, \prl, \href
  {https://ui.adsabs.harvard.edu/abs/1999PhRvL..82.2832S} {82, 2832}

\bibitem[\protect\citeauthoryear{{Vegetti}, {Despali}, {Lovell}  \&
  {Enzi}}{{Vegetti} et~al.}{2018}]{2018MNRAS.481.3661V}
{Vegetti} S.,  {Despali} G.,  {Lovell} M.~R.,   {Enzi} W.,  2018, \mn@doi
  [\mnras] {10.1093/mnras/sty2393}, \href
  {https://ui.adsabs.harvard.edu/abs/2018MNRAS.481.3661V} {481, 3661}

\bibitem[\protect\citeauthoryear{Viel, Becker, Bolton  \& Haehnelt}{Viel
  et~al.}{2013}]{Viel:2013fqw}
Viel M.,  Becker G.~D.,  Bolton J.~S.,   Haehnelt M.~G.,  2013, \mn@doi [Phys.
  Rev. D] {10.1103/PhysRevD.88.043502}, 88, 043502

\bibitem[\protect\citeauthoryear{Vogel \& Abazajian}{Vogel \&
  Abazajian}{2022}]{Vogel:2022odl}
Vogel C.~M.,  Abazajian K.~N.,  2022, arXiv e-prints

\bibitem[\protect\citeauthoryear{Vogt, Marsh  \& Lagu\"e}{Vogt
  et~al.}{2022}]{Vogt:2022bwy}
Vogt S. M.~L.,  Marsh D. J.~E.,   Lagu\"e A.,  2022, arXiv e-prints

\bibitem[\protect\citeauthoryear{{Wizinowich} et~al.,}{{Wizinowich}
  et~al.}{2022}]{KAPA}
{Wizinowich} P.,  et~al., 2022, in {Schreiber} L.,  {Schmidt} D.,   {Vernet}
  E.,  eds,  Society of Photo-Optical Instrumentation Engineers (SPIE)
  Conference Series Vol. 12185, Adaptive Optics Systems VIII. p. 121850Q,
  \mn@doi{10.1117/12.2628275}

\bibitem[\protect\citeauthoryear{{Wright} et~al.,}{{Wright}
  et~al.}{2019}]{LIGER}
{Wright} S.,  et~al., 2019, in Bulletin of the American Astronomical Society.
  p.~201

\bibitem[\protect\citeauthoryear{Zelko, Treu, Abazajian, Gilman, Benson,
  Birrer, Nierenberg  \& Kusenko}{Zelko et~al.}{2022}]{Zelko:2022tgf}
Zelko I.~A.,  Treu T.,  Abazajian K.~N.,  Gilman D.,  Benson A.~J.,  Birrer S.,
   Nierenberg A.~M.,   Kusenko A.,  2022, \mn@doi [Phys. Rev. Lett.]
  {10.1103/PhysRevLett.129.191301}, 129, 191301

\makeatother
\end{thebibliography}

%\begin{thebibliography}{99}
%\end{thebibliography}

%%%%%%%%%%%%%%%%%%%%%%%%%%%%%%%%%%%%%%%%%%%%%%%%%%

%%%%%%%%%%%%%%%%% APPENDICES %%%%%%%%%%%%%%%%%%%%%

%\appendix


%%%%%%%%%%%%%%%%%%%%%%%%%%%%%%%%%%%%%%%%%%%%%%%%%%

% Don't change these lines
\bsp	% typesetting comment
\label{lastpage}
\end{document}